\documentclass[twocolumn,aps,pra,showpacs]{revtex4}

\usepackage{graphicx}

\begin{document}
\title{Bell's inequalities with realistic noise for polarization-entangled photons}
\author{Ad\'{a}n Cabello}
\email{adan@us.es}
\affiliation{Departamento de F\'{\i}sica Aplicada II,
Universidad de Sevilla, 41012 Sevilla, Spain}
\author{\'{A}lvaro Feito}
\email{alvaro.feito@gmail.com}
\affiliation{Departamento de F\'{\i}sica Aplicada II,
Universidad de Sevilla, 41012 Sevilla, Spain}
\author{Ant\'{\i}a Lamas-Linares}
\email{antia_lamas@nus.edu.sg}
\affiliation{Quantum Information Technology Lab, Physics Department, National University of Singapore,\\
2 Science Drive 3, 117542 Singapore}
\date{\today}


\begin{abstract}
Contrary to the usual assumption that the experimental preparation
of pure entangled states can be described by mixed states due to
white noise, a more realistic description for
polarization-entangled states produced by parametric
down-conversion is that they are mixed states due to decoherence
in a preferred polarization basis. This distinction between white
and colored noise is crucial when we look for maximal violations
of Bell's inequalities for two-qubit and two-qutrit entangled
states. We find that violations of Bell's inequalities with
realistic noise for polarization-entangled photons are extremely
robust for colored noise, whereas this is not the case for white
noise. In addition, we study the difference between white and
colored noise for maximal violations of Bell's inequalities for
three and four-qubit entangled states.
\end{abstract}


\maketitle


\section{Introduction}


Entanglement is an essential resource for quantum information
processing~\cite{NC00} and the violation of Bell's
inequalities~\cite{Bell64} can be a basic tool test to detect
entanglement~\cite{Terhal98} and discard the possibility of
simulating the experimental data by means of classically
correlated systems. So far, the most reliable source of both
two-party and multiparty entanglement are polarization-entangled
photons created by parametric down-conversion
(PDC)~\cite{KMWZSS95}. Although there are other types of photon
entanglement (momentum entanglement~\cite{RT90}, position and time
entanglement~\cite{Franson89}, time-bin
entanglement~\cite{BGTZ99}, orbital angular momentum
entanglement~\cite{MVWZ01}), and even entanglement between
ions~\cite{RKMSIMW01}, or between atomic beams~\cite{VXK02},
polarization entanglement remains the most widely implemented due
to its robustness and ease of use.

When violations of Bell's inequalities for realistic states are
analyzed and, specifically, when resistance to noise is studied,
it is usually assumed that such a noise is white or
uncolored~\cite{KGZMZ00,DKZ01,CKKOZ01,KKCZO02,CGLMP02}. In the
presence of white noise a quantum pure state $|\psi\rangle$
becomes
\begin{equation}
\rho = p\,|\psi\rangle \langle \psi|+ \frac{1-p}{d}
1\!\!\:\!{\rm{I}}, \label{Uncoloredstates}
\end{equation}
where $p$ is the probability that the state is unaffected by
noise, $d$ is the dimension of the Hilbert space of the whole
system, and $1\!\!\:\!{\rm{I}}$ is the identity matrix in that
Hilbert space.

However, when working with real systems the noise is very rarely
colorless. For entangled photon production via PDC it is
experimentally found that while correlations are very strong in
the ``natural'' basis of the crystal (i.e., the basis in which the
phase matching conditions are expressed), the same is not true for
maximally conjugated basis. For this reason it is common to use
the visibility obtained by fixing one polarizer at $45^{\circ}$
and rotating the other as a shorthand for entanglement quality.
The physical reason for the difference in the two visibilities
lies in the phase matching. For type-II PDC every down-converted
pair will consist of one ordinary and one extraordinary photon
(generally labeled as $H$ and $V$), but by itself the phase
matching does not guarantee correlations in any other basis. To
achieve an entangled state, it is necessary to make the two
down-conversion possibilities, $|HV\rangle$ and $|VH\rangle$,
indistinguishable and obtain a fixed phase between them. This
indistinguishability is achieved by careful mode selection and
enhanced by the use of narrow band filters and so-called
compensator crystals~\cite{KMWZSS95} but it is inevitably
imperfect. Since in an experiment we typically only measure the
polarization, the other degrees of freedom are traced over and the
noise appears as a decoherence-type term
\begin{equation}
\rho = p\,|\psi\rangle \langle \psi|+ \frac{1-p}{2}( |HV\rangle
\langle HV|+|VH\rangle \langle VH|).
\end{equation}

In this paper, we study the robustness of maximal violations of
several Bell's inequalities against two kinds of noise: white
noise and the colored noise mentioned above. The experimental
scenarios studied correspond to experiments already performed in
the laboratory, and therefore the conclusions can be checked using
current technology. In Sec.~\ref{sec:2} we study the influence of
noise on the maximal violation of the Clauser-Horne-Shimony-Holt
(CHSH) inequality~\cite{CHSH69} by two-qubit entangled states,
assuming that these states are maximally entangled states affected
by either of these types of noise. In Sec.~\ref{sec:3} we study
the dependence of the maximal violation of the Mermin-Klyshko
inequalities~\cite{Mermin90,Klyshko93,BK93} by three-qubit and
four-qubit entangled states, assuming that these states are
Greenberger-Horne-Zeilinger (GHZ) states~\cite{GHZ89} affected by
white or colored noise. In Sec.~\ref{sec:4} we study the influence
of noise on the maximal violation of a CHSH-like inequality by
two-qutrits entangled states simulated by four-qubit entangled
states, assuming that these states are maximally entangled
two-qutrit states simulated by rotationally invariant four-qubit
singlet states~\cite{LHB01,HLB02,EGBKZW04,BEGKCW04,BEKGWGHBLA04}
affected by white or colored noise. Finally, the conclusions are
summarized in Sec.~\ref{sec:5}.


\section{Two-photon states and the CHSH inequality}
\label{sec:2}


For two qubits, singlet states,
\begin{equation}
|\psi^-\rangle = \frac{1}{\sqrt{2}} \left(|01\rangle-|10\rangle
\right),
\end{equation}
affected by white noise,
\begin{equation}
\rho_{\rm W} = p\,|\psi^-\rangle \langle \psi^-|+ \frac{1-p}{4}
1\!\!\:\!{\rm{I}}, \label{Wernerstates}
\end{equation}
are called Werner states~\cite{Werner89}. It is usually assumed
that Werner states suitably describe the states employed in
two-qubit tests of the Bell's inequality using
polarization-entangled photons prepared using
PDC~\cite{WJSWZ98,BCDC04}.

Experimental evidence and physical arguments show that a colorless
noise model is not the best choice for describing states produced in
type-II PDC. A more realistic description is given by an alternative
one parameter model where the singlet is mixed with decoherence
terms in a preferred polarization basis
\begin{equation}
\rho_{\rm C} = p\,|\psi^-\rangle \langle \psi^-|+ \frac{1-p}{2}
(|01\rangle\langle 01|+|10\rangle\langle 10|).
\label{Antiastates}
\end{equation}

The CHSH inequality~\cite{CHSH69} is
\begin{equation}
|\beta| \le 2, \label{CHSH}
\end{equation}
where
\begin{equation}
\beta = -\langle A_0 B_0\rangle - \langle A_0 B_1\rangle -
\langle A_1 B_0\rangle + \langle A_1 B_1\rangle,
\label{Belloperator}
\end{equation}
is called the Bell operator.

To study maximal violations of the CHSH inequality~(\ref{CHSH})
for states with white noise $\rho_{\rm W}$, given by
Eq.~(\ref{Wernerstates}), and colored noise $\rho_{\rm C}$, given
by Eq.~(\ref{Antiastates}), it is sufficient to consider the
following one-qubit observables:
\begin{eqnarray}
A_0 & = & \sigma_z,
\label{A0} \\
A_1 & = & \cos(\theta) \sigma_z+\sin(\theta) \sigma_x, \\
B_0 & = & \cos(\phi) \sigma_z+\sin(\phi) \sigma_x, \\
B_1 & = & \cos(\phi-\theta) \sigma_z+\sin(\phi-\theta) \sigma_x,
\label{B1}
\end{eqnarray}
where $\sigma_z$ and $\sigma_x$ are the usual Pauli matrices. Our
aim is to study the dependence on $p$ of the maximum violation of
the CHSH inequality (\ref{CHSH}).

The Bell operator (\ref{Belloperator}) for the $\rho_{\rm W}$
states and the local observables (\ref{A0})--(\ref{B1}) is
\begin{eqnarray}
\beta_{\rm W}(p, \theta, \phi) & = & 2 p \left\{ \cos(\phi)
\left[\sin^2(\theta)+\cos(\theta)\right] \right. \nonumber \\
& & \left. -\sin(\phi) \left[\cos(\theta)-1\right] \sin(\theta)
\right\}.
\end{eqnarray}
The Bell operator for the $\rho_{\rm C}$ states and the local
observables (\ref{A0})--(\ref{B1}) is
\begin{eqnarray}
\beta_{\rm C}(p, \theta, \phi) & = & \cos(\phi) \left[(1 + p)
\sin^2(\theta)+2 \cos(\theta)\right] \nonumber \\ & & - \sin(\phi)
(1 + p) \left[\cos(\theta)-1\right] \sin(\theta).
\end{eqnarray}

For the $\rho_{\rm W}$ states, the maximum possible value of
$\beta$, as a function of $p$, is
\begin{equation}
\beta_{\rm W max} (p) = 2 \sqrt{2} p.
\label{BetaWerner}
\end{equation}
Therefore, Werner states only violate the CHSH inequality if $p >
1/\sqrt{2} \approx 0.707$. However, $\rho_{\rm W}$ states are
entangled if $p \ge \frac{1}{3}$~\cite{Peres96}. For any~$p$, the maximum value of $\beta$
is always obtained by choosing
\begin{eqnarray}
\theta & = & \frac{\pi}{2},
\label{theta} \\
\phi & = & \frac{\pi}{4}.
\label{phi}
\end{eqnarray}

If one insists on performing a test of the CHSH inequality with
$\rho_{\rm C}$ states, but using the angles~(\ref{theta}) and
(\ref{phi}), then the maximum $\beta$ is given by
\begin{eqnarray}
\beta_{\rm C} (p, \theta=\pi /2, \phi=\pi/4) = \sqrt{2} (1+p).
\end{eqnarray}
Therefore, in this case there would no longer be a violation of
the CHSH inequalities for $p \le \sqrt{2}-1 \approx 0.41$.

\begin{figure}
\centerline{\includegraphics[trim=20 20 10
20,clip=true,angle=-90,width=10.0cm]{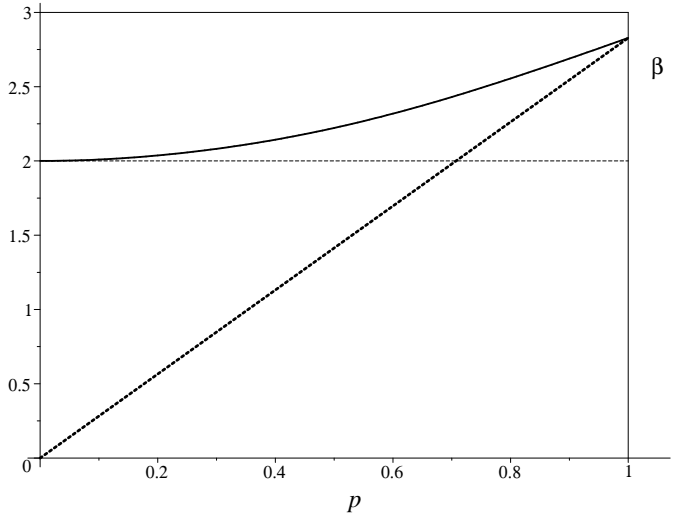}}
\caption{\label{Beta2} Maximum possible violation of the CHSH
inequality as a function of $p$ for states with colored noise
$\rho_{\rm C}$ given by Eq.~(\ref{Antiastates}) (upper continuous
line), and states with white noise $\rho_{\rm W}$ given by
Eq.~(\ref{Wernerstates}) (lower dashed straight line). The
classical bound is $2$ and the maximal violation, for $p=1$, is $2
\sqrt{2} \approx 2.83$.}
\end{figure}

\begin{figure}
\centerline{\includegraphics[trim=20 20 10
20,clip=true,angle=-90,width=10.0cm]{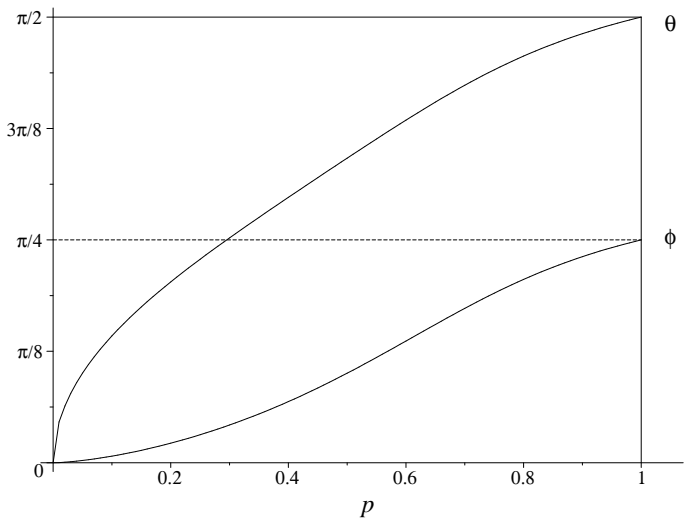}}
\caption{\label{Phi2} Optimal values of the local parameters
$\theta$ and $\phi$ giving the maximum violation of the CHSH
inequality for the $\rho_{\rm C}$ states as a function of $p$.}
\end{figure}

However, for the $\rho_{\rm C}$ states, the maximum violation of
the CHSH inequality (i.e., the maximum value of $\beta$) depends
on $p$ in a more complicated fashion. The dependence of the
maximum value of $\beta$ with $p$ for the $\rho_{\rm C}$ states is
illustrated in Fig.~\ref{Beta2}. Indeed, for different $p$ these
maximum violations occur for different values of the angles
$\theta$ and $\phi$, as illustrated in Fig.~\ref{Phi2}.


The first interesting point is that the $\rho_{\rm C}$ states
violate the CHSH inequality {\em for any $p$}. That is, the
violation is very robust against the colored noise. On the other
hand, these maximum violations occur for local observables which
depend on $p$.


\section{Three and four-qubit GHZ states and Mermin-Klyshko inequalities}
\label{sec:3}


Consider the three-qubit version of Mermin's
inequality~\cite{Mermin90}
\begin{equation}
|\mu| \le 2, \label{M3}
\end{equation}
where
\begin{equation}
\mu = \langle A_1 B_2 B_3\rangle + \langle B_1 A_2 B_3\rangle +
\langle B_1 B_2 A_3\rangle - \langle A_1 A_2 A_3\rangle.
\label{Merminoperator}
\end{equation}
It is usually assumed that GHZ states with white noise suitably
describe the states employed in three-qubit tests of the Bell
inequality using polarization-entangled
photons~\cite{BPDWZ99,PBDWZ00}. These states can be written as
\begin{equation}
\rho_{\rm W} = p\,|{\rm GHZ_3}\rangle \langle {\rm GHZ_3}|+
\frac{1-p}{8} 1\!\!\:\!{\rm{I}}, \label{GHZWhite}
\end{equation}
where
\begin{equation}
|{\rm GHZ_3}\rangle = \frac{1}{\sqrt{2}}
\left(|000\rangle-|111\rangle \right). \label{GHZ}
\end{equation}
However, a more realistic descriptions of the states obtained in
the laboratory is
\begin{eqnarray}
\rho_{\rm C} & = & p\,|{\rm GHZ_3}\rangle \langle {\rm GHZ_3}| \nonumber \\
& & + \frac{1-p}{2} \left(|000\rangle \langle 000|+|111\rangle
\langle 111|\right). \label{GHZLamas}
\end{eqnarray}

To calculate the maximum value of $\mu$ for the $\rho_{\rm W}$ and
$\rho_{\rm C}$ states we will confine our attention to the
following local observables:
\begin{eqnarray}
A_j & = & \cos(\theta) \sigma_x+\sin(\theta) \sigma_y, \label{Aj} \\
B_j & = & \cos(\phi) \sigma_x+\sin(\phi) \sigma_y. \label{Bj}
\end{eqnarray}

Then, {\em both} for states with white (\ref{GHZWhite}) and
colored noise (\ref{GHZLamas}), we obtain that
\begin{equation}
\mu_{\rm max} (p) = 4 p.
\label{MuMax}
\end{equation}
For any $p$, these maximum values can be obtained by choosing
\begin{eqnarray}
\theta & = & 0,
\label{theta3} \\
\phi & = & \pi / 2,
\label{phi3}
\end{eqnarray}
in Eqs.~(\ref{Aj}) and (\ref{Bj}). Therefore, in this case, the
more realistic noise does not give a different violation than
those with white noise.

The conclusion is similar for four-photon GHZ states and
Mermin-Klyshko inequality~\cite{Mermin90,Klyshko93,BK93}:
\begin{equation}
| \kappa | \le 4, \label{K4}
\end{equation}
where
\begin{eqnarray}
\kappa & = & \langle B_1 A_2 A_3 A_4\rangle+\langle A_1 B_2 A_3 A_4\rangle+\langle A_1 A_2 B_3 A_4\rangle \nonumber \\
& & +\langle A_1 A_2 A_3 B_4\rangle +\langle B_1 A_2 A_3 B_4\rangle +\langle A_1 B_2 A_3 B_4\rangle \nonumber \\
& & +\langle A_1 A_2 B_3 B_4\rangle -\langle B_1 B_2 B_3 B_4\rangle +\langle A_1 B_2 B_3 A_4\rangle \nonumber \\
& & +\langle B_1 A_2 B_3 A_4\rangle +\langle B_1 B_2 A_3 A_4\rangle -\langle A_1 A_2 A_3 A_4\rangle \nonumber \\
& & -\langle A_1 B_2 B_3 B_4\rangle -\langle B_1 A_2 B_3 B_4\rangle -\langle B_1 B_2 A_3 B_4\rangle \nonumber \\
& & -\langle B_1 B_2 B_3 A_4\rangle. \label{MKoperator}
\end{eqnarray}

Both for the states with white noise
\begin{equation}
\rho_{\rm W} = p\,|{\rm GHZ_4}\rangle \langle {\rm GHZ_4}|+
\frac{1-p}{16} 1\!\!\:\!{\rm{I}}, \label{GHZ4White}
\end{equation}
and for the states with colored noise
\begin{eqnarray}
\rho_{\rm C} & = & p\,|{\rm GHZ_4}\rangle \langle {\rm GHZ_4}| \nonumber \\
& & +\frac{1-p}{2} \left(|0000\rangle \langle 0000|+|1111\rangle
\langle 1111|\right), \label{GHZ4Lamas}
\end{eqnarray}
where
\begin{equation}
|{\rm GHZ_4}\rangle = \frac{1}{\sqrt{2}}
\left(|0000\rangle-|1111\rangle \right), \label{GHZ4}
\end{equation}
and choosing the same six local observables for the first three
qubits given by Eqs.~(\ref{Aj}) and (\ref{Bj}), plus the following
local observables on the fourth qubit:
\begin{eqnarray}
A_4 & = & \cos(\theta) \frac{\sigma_{x}+\sigma_{y}}{\sqrt{2}} + \sin(\theta) \frac{\sigma_{y}-\sigma_{x}}{\sqrt{2}},\\
B_4 & = & \cos(\phi)\frac{\sigma_{x}+\sigma_{y}}{\sqrt{2}} +
\sin(\phi)\frac{\sigma_{y}-\sigma_{x}}{\sqrt{2}},
\end{eqnarray}
we find that $|\kappa|= p f(\theta,\phi)$, where
$f_{\rm{max}}=8\sqrt{2}$. Therefore, $|\kappa|$ will attain a
maximum for a given $p$ at fixed angles, which turn out to be
Eqs.~(\ref{theta3}) and (\ref{phi3}).

Since inequalities (\ref{M3}) and (\ref{K4}) are tools to detect
and measure genuine $N$-particle
nonseparability~\cite{SU02,CGPRS02,SS02,Cereceda02}, we conclude
that there is no difference between the white and colored noise's
entanglement for an experiment with the proposed observables with
three~\cite{BPDWZ99,PBDWZ00} or four~\cite{PDGWZ01,ZYCZZP03}
polarization-entangled GHZ states .


\section{Two-qutrit singlet state simulated with four photons and a CHSH-like inequality}
\label{sec:4}


The two-qutrit singlet state (i.e., the two spin-1 particles'
singlet state),
\begin{equation}
|\psi_{3 \times 3}\rangle = \frac{1}{\sqrt{3}}\left(|-1,+1\rangle
- |0,0\rangle + |+1,-1\rangle\right) \label{psi3}
\end{equation}
can be simulated by a four-qubit state by defining
\begin{eqnarray}
|-1\rangle & := & |00\rangle, \\
|0\rangle & := & \frac{1}{\sqrt{2}}\left(|01\rangle+|10\rangle\right), \\
|+1\rangle & := & |11\rangle.
\end{eqnarray}
Substituting in Eq.~(\ref{psi3}), we obtain
\begin{eqnarray}
|\psi_{4 \times 4} \rangle & = & {1 \over 2 \sqrt{3}}
(2|0011\rangle-|0101\rangle-|0110\rangle \nonumber \\
& & -|1001\rangle-|1010\rangle+2|1100\rangle),
\label{psi4}
\end{eqnarray} which is the four-qubit rotationally
invariant state which can be prepared in the
laboratory~\cite{LHB01,HLB02,EGBKZW04,BEGKCW04,BEKGWGHBLA04}.


\begin{figure}
\centerline{\includegraphics[trim=20 20 10
20,clip=true,angle=-90,width=10.0cm]{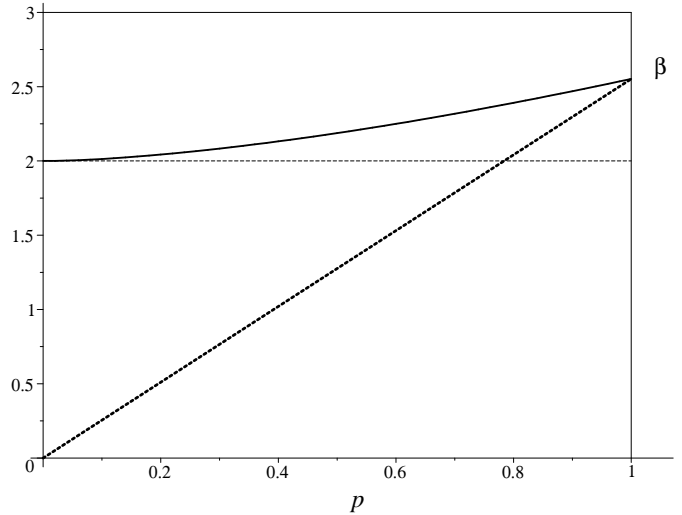}}
\caption{\label{Beta4} Maximum violation of the CHSH-like
inequality given by Eq.~(\ref{CHSH4}) as a function of $p$ for
states with colored noise $\rho_{\rm C}$, given by
Eq.~(\ref{psi4colored}) (upper bold line), and states with white
noise $\rho_{\rm W}$, given by Eq.~(\ref{psi4white}) (lower dashed
straight line). The classical bound is $2$ and the maximal
violation, for $p=1$, is $2 (1+2 \sqrt{2})/3 \approx 2.55$.}
\end{figure}


\begin{figure}
\centerline{\includegraphics[trim=20 20 10
20,clip=true,angle=-90,width=10.0cm]{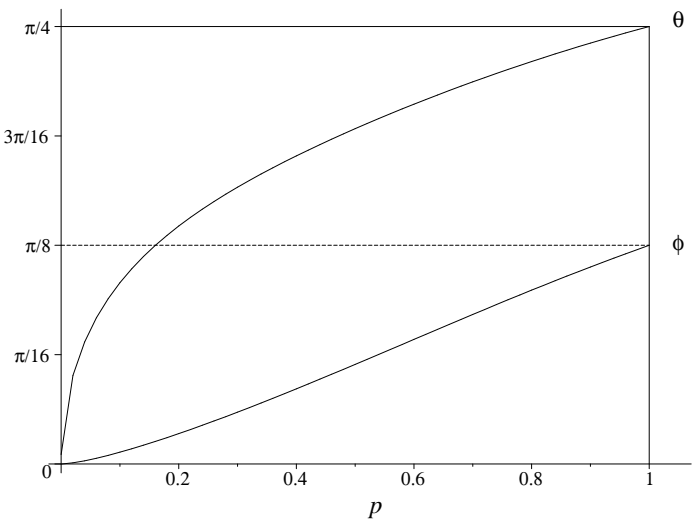}}
\caption{\label{Phi4} Optimal values of the local parameters
$\theta$ (up) and $\phi$ (down) giving the maximum violation of
the CHSH-like inequality (\ref{CHSH4}) for states with colored
noise (\ref{psi4colored}) as a function of $p$.}
\end{figure}


We will consider two types of noise: white noise and colored
noise. The four-qubit rotationally invariant state with white
noise is
\begin{eqnarray}
\rho_{\rm W} & = & p\,|\psi_{4 \times 4}\rangle \langle \psi_{4
\times 4}|+ \frac{1-p}{16} 1\!\!\:\!{\rm{I}}, \label{psi4white}
\end{eqnarray}
where $1\!\!\:\!{\rm{I}}$ is the identity matrix. The four-qubit
rotationally invariant state with colored noise is
\begin{eqnarray}
\rho_{\rm C} & = & p\,|\psi_{4 \times 4}\rangle \langle \psi_{4 \times 4}| \nonumber \\
& & +\frac{1-p}{12}
(4|0011\rangle \langle 0011|+|0101\rangle \langle 0101| \nonumber \\
& & +|0110\rangle \langle 0110|+|1001\rangle \langle 1001| \nonumber \\
& & +|1010\rangle \langle 1010|+4|1100\rangle \langle 1100|),
\label{psi4colored}
\end{eqnarray}

We want to study how the maximum possible violation of a CHSH-like
inequality for the two-qutrit singlet state simulated by a
four-qubit rotationally invariant state depends on noise.
Simulating state (\ref{psi3}) by means of state (\ref{psi4})
imposes some restrictions over the possible local observables. The
first is that we must consider only those which can be implemented
by the product of two one-photon observables, since general
two-photon observables are difficult to implement. The second is
that some procedures for preparing state (\ref{psi4}) do not allow
us to distinguish between photon-1 and photon-2 (and between
photon-3 and photon-4); therefore, this will lead us to consider
only local measurements that can be implemented as a product of
the {\em same} two one-photon polarization observables. Therefore,
we will study violations of the following CHSH-like inequality
\begin{equation}
|\beta| \le 2, \label{CHSH4}
\end{equation}
where
\begin{eqnarray}
\beta & = & \langle A_0 A_0 B_0 B_0\rangle+\langle A_0 A_0 B_1
B_1\rangle+\langle A_1 A_1 B_0 B_0\rangle \nonumber \\ & &
-\langle A_1 A_1 B_1 B_1\rangle,
\label{Belloperator2}
\end{eqnarray}
where, for instance, $A_0 A_0 B_0 B_0$ is the product of the
results of measuring $A_0$ on photons 1 and 2, and $B_0$ on
photons 3 and 4.

To study maximal violations of Bell's inequality (\ref{CHSH4}) for
states (\ref{psi4white}) and (\ref{psi4colored}), it is sufficient
to consider the one-photon observables (\ref{A0})--(\ref{B1}).
Then, the Bell operator (\ref{Belloperator2}) for states with
white noise (\ref{psi4white}) is given by
\begin{eqnarray}
\beta_{\rm W}(p, \theta, \phi) & = & \frac{2p}{3} \left\{ 2 \left[
\cos \left( 2 \theta - 2 \phi \right) + \cos^2\left(\phi\right) \right] \right. \nonumber \\
& & \left. - \cos \left[4 \theta - 2 \phi\right]\right\}.
\end{eqnarray}
The Bell operator for the states with colored noise (\ref{psi4colored}) is given by
\begin{eqnarray}
\beta_{\rm C}(p, \theta, \phi) & = & \frac{1}{24}
\left[24-8 p- \left(3+13 p\right) \cos\left(4 \theta-2 \phi\right) \right. \nonumber \\
& & +\left(9+23 p\right) \cos\left(2 \theta-2 \phi\right)+\left(15+p\right) \cos\left(2 \phi\right) \nonumber \\
& & \left. +\left(3-3p \right) \cos\left(2 \theta+2 \phi\right)\right].
\end{eqnarray}

For states with white noise (\ref{psi4white}), the maximum possible value of
$\beta$ as a function of $p$ is
\begin{equation}
\beta_{\rm W max} (p) = \frac{2}{3} \left( 1 + 2 \sqrt{2} \right) p.
\label{BetaWhite}
\end{equation}
Therefore, states with white noise (\ref{psi4white}) only violate
inequality (\ref{CHSH4}) if $p > 3/(1+2\sqrt{2}) \approx 0.784$.
For any~$p$, the maximum value of $\beta$ is always obtained by
choosing
\begin{eqnarray}
\theta & = & \frac{\pi}{4}, \\
\phi & = & \frac{\pi}{8}.
\end{eqnarray}

However, for states with colored noise (\ref{psi4colored}), the
maximum value of $\beta$ depends on $p$ in a more complicated
fashion. The dependence of the maximum value of $\beta$ on $p$ for
states with colored noise (\ref{psi4colored}) is illustrated in
Fig.~\ref{Beta4}. These maximum violations occur for different
values of the angles $\theta$ and $\phi$, depending on the value
of $p$, as illustrated in Fig.~\ref{Phi4}.

A remarkable property is that for $p=0$, and choosing
$\theta=\phi=0$, we obtain $\beta = 2$. Moreover, violations of
inequality (\ref{CHSH4}) occur {\em for any $p>0$}. That is, the
violation is extremely robust against noise.


\section{Conclusions}
\label{sec:5}


Polarization-entangled photons created by PDC are, so far, the most
reliable and widespread systems to prepare the most common types of
entanglement: two-qubit entanglement, multiqubit entanglement, and
two-qudit entanglement. Testing the violation of Bell's inequalities
is a basic tool for detecting entanglement and, therefore, for
confirming the genuine quantum behavior of a physical system. Real
observed data are affected by noise and most theoretical studies of
violations of Bell's inequalities assume that this noise is white.
However, PDC sources have their own characteristic noise. In this
paper we have investigated to what extent this specific noise
modifies previous conclusions about the influence of noise in Bell's
inequalities based on the assumption that the noise is completely
unbiased. The most important conclusion is that, in the case of
bipartite systems of qubits or qutrits (each of them simulated by a
pair of qubits), the violation of the CHSH inequality is extremely
robust, meaning that even sources with an extremely low purity $p$
violate the CHSH inequality with a suitable choice of local
observables. We have calculated which local observables provide the
maximum violation as a function of $p$ for both cases. Not
surprisingly we find that the case of white noise, the maximal
violation always occurs for the same choice of local observables and
no violation occurs under a certain value of $p$, while in the more
realistic case the optimal observables are a function of $p$.

Our predictions for both types of noise and different values of
$p$ can be experimentally tested in the laboratory with current
technology, since there are sources of high purity ($p > 0.98$),
and it is possible to generate additional noise of both types with
relative ease~\cite{LPLK05}.
We have also studied the influence of realistic noise for three
and four-qubit systems and Mermin-Klyshko inequalities and we have
found that, in that instance, the possible maximal violations are
similar to those obtained assuming white noise.


\section*{Acknowledgments}


The authors thank A. Ac\'{\i}n and M.~\.{Z}ukowski for useful
discussions. A.C. acknowledges support from A*STAR Grant
No.~R-144-000-071-089-112 during his visit to the National
University of Singapore, and from projects No.~BFM2002-02815 and
No.~FQM-239. A.L.-L. acknowledges support from A*STAR Grant
No.~R-144-000-071-089-112.



\end{document}